\documentclass[%
 aip,
 amsmath,amssymb,
 reprint,%
]{revtex4-1}

\usepackage{graphicx}
\usepackage{dcolumn}
\usepackage{bm}

\usepackage[utf8]{inputenc}
\usepackage[T1]{fontenc}
\usepackage{mathptmx,microtype}
\usepackage{etoolbox}
\usepackage{xcolor}

\newcommand{\sfo}{SrFeO$_3$}
\newcommand{\sfox}{SrFeO$_x$}
\newcommand{\sfoil}{SrFeO$_2$}
\newcommand{\sfod}{SrFeO$_{3-\delta}$}
\newcommand{\sfobm}{SrFeO$_{2.5}$}
\newcommand{\sto}{SrTiO$_3$}
\newcommand{\alumina}{Al$_2$O$_3$}

\makeatletter
\def\@email#1#2{%
 \endgroup
 \patchcmd{\titleblock@produce}
  {\frontmatter@RRAPformat}
  {\frontmatter@RRAPformat{\produce@RRAP{*#1\href{mailto:#2}{#2}}}\frontmatter@RRAPformat}
  {}{}
}%
\makeatother
\begin{document}

\preprint{AIP/123-QED}

\title[Designing heterostructures to control oxygen stoichiometry in helimagnetic perovskite strontium ferrite]{Designing heterostructures to control oxygen stoichiometry in helimagnetic perovskite strontium ferrite}
\author{J. Fowlie}

\affiliation{ 
Stanford Institute for Materials and Energy Sciences, SLAC National Accelerator Laboratory, Menlo Park, CA 94025, USA}

\affiliation{ 
Department of Applied Physics, Stanford University, Stanford, CA 94305, USA
}%
\affiliation{Department of Materials Science and Engineering, Northwestern University, Evanston, 60208, USA}

  \email{jfowlie@northwestern.edu.}

\author{B. Mundet}%

\affiliation{ 
Catalan Institute of Nanoscience and Nanotechnology (ICN2), Barcelona 08193 Catalonia, Spain
}%

\author{D. Puggioni}
\affiliation{Department of Materials Science and Engineering, Northwestern University, Evanston, 60208, USA}

\author{L. Bhatt}
\affiliation{School of Applied and Engineering Physics, Cornell University, Ithaca, New York, 14850, USA}

\author{E. R. Hoglund}
\affiliation{Center for Nanophase Materials Sciences, Oak Ridge National Laboratory, Oak Ridge, TN, 37830, USA}

\author{W. J. Kim}
\affiliation{ 
Stanford Institute for Materials and Energy Sciences, SLAC National Accelerator Laboratory, Menlo Park, CA 94025, USA
}%
\affiliation{ 
Department of Applied Physics, Stanford University, Stanford, CA 94305, USA
}%
\affiliation{Department of Material Science Engineering, Pusan National University, Busan 46241, Republic of Korea}

\author{J. Li}
\affiliation{ 
Stanford Institute for Materials and Energy Sciences, SLAC National Accelerator Laboratory, Menlo Park, CA 94025, USA
}%
\affiliation{ 
Department of Applied Physics, Stanford University, Stanford, CA 94305, USA
}%

\author{S. J. Lee}
\affiliation{ 
Stanford Synchrotron Radiation Lightsource, SLAC National Accelerator Laboratory, Menlo Park, CA 94025, USA
}%

\author{W. Liu}
\affiliation{Department of Materials Science and Engineering, Northwestern University, Evanston, 60208, USA}

\author{A. Devincenti}
\affiliation{Department of Materials Science and Engineering, Northwestern University, Evanston, 60208, USA}

\author{J. M. Rondinelli}
\affiliation{Department of Materials Science and Engineering, Northwestern University, Evanston, 60208, USA}

\author{D. A. Muller}
\affiliation{School of Applied and Engineering Physics, Cornell University, Ithaca, New York, 14850, USA}
\affiliation{Kavli Institute at Cornell for Nanoscale Technology, Cornell University, Ithaca, New York, 14850, USA}

\author{H. Y. Hwang}
\affiliation{ 
Stanford Institute for Materials and Energy Sciences, SLAC National Accelerator Laboratory, Menlo Park, CA 94025, USA
}%
\affiliation{ 
Department of Applied Physics, Stanford University, Stanford, CA 94305, USA
}%

\date{\today}

\begin{abstract}
A large challenge in determining the physics of helimagnetic \sfo{} is in stabilizing the stoichiometric chemical phase over long enough time scales to conduct extensive measurements. Degradation in \sfo{} manifests mainly as a crossover from metallic to insulating behavior. Using a combination of electronic transport and density functional theory, we show that this degradation is dominated by oxygen loss, possibly on the order of one percent. We further demonstrate that high quality \sfo{} thin films can be stabilized long-term by combining a nanoscale band insulator capping layer with an \textit{ex situ} ozone anneal. We show that this produces a nearly-pristine cation sublattice and preserves metallicity for at least several weeks.
These results establish a reliable pathway for producing chemically stable \sfo{} thin films, enabling reproducible studies of its unusual helimagnetism.
\end{abstract}

\maketitle

\section{\label{sec:level1}Introduction}

Perovskite \sfo{} is an unusual example of a helimagnetic compound that is centrosymmetric and cubic, so its helimagnetism cannot be explained by Dzyaloshinskii-Moriya physics or by frustration \cite{Takeda1972,Ishiwata2011,Ishiwata2020}. Helimagnetism, and topologically-protected magnetic structures that are reported in this compound \cite{Ishiwata2020}, also make \sfo{} an interesting material in the context of spintronic, magnonic, qubit, and high-density memory applications \cite{Islam2023,Psaroudaki2023}.  

One of the primary impediments to the further study of helimagnetism in \sfo{} is that the compound is metastable at ambient conditions. This metastability is believed to be driven by the relatively unfavorable Fe$^{4+}$ valence state and its tendency to reduce toward Fe$^{3+}$. This leads to two challenges: (1) as-grown \sfo{} is usually reported to be oxygen deficient \sfod{} \cite{Takeda1986,Schmidt2000}, and (2) \sfo{} (or \sfod{}) will rapidly degrade (further), preventing reliable characterization, especially multi-probe, at large facilities, or over an extended period of time \cite{Wang2020e, Enriquez2016}.

Although a hindrance to studies of electronic and magnetic properties, this evolution in oxygen stoichiometry has brought much recent attention to \sfo{} because it facilitates superior oxygen-ion transport properties for potential application in nonvolatile neuromorphic hardware \cite{Nallagatla2019, Ge2019}. In particular, electric field is found to drive a reversible phase transition between the perovskite phase and the brownmillerite \sfobm{} \cite{Ge2019}. Such a transition involves only the movement of oxygen anions and can be considered as a topotactic redox reaction, where topotactic refers to the fact that the arrangement of the cation sublattice remains the same.

Precise control of the oxidation state of \sfox{} and, in particular, preventing the fully oxidized \sfo{} from degrading, has yet to be demonstrated.
Here we report the positive effect of a nanometric band insulator capping layer on the transport and microstructure properties of \sfox{} thin films. Such a capping layer produces the highest quality \sfo{} reported to date, provides evidence that the nature of the degradation is indeed oxygen loss and is reversible, and stabilizes high quality \sfo{} over many weeks, enabling extensive investigation of electronic and magnetic properties \cite{Fowlie2026}.

\section{Results}
\subsubsection{Growth and characterization}

\begin{figure}
\includegraphics{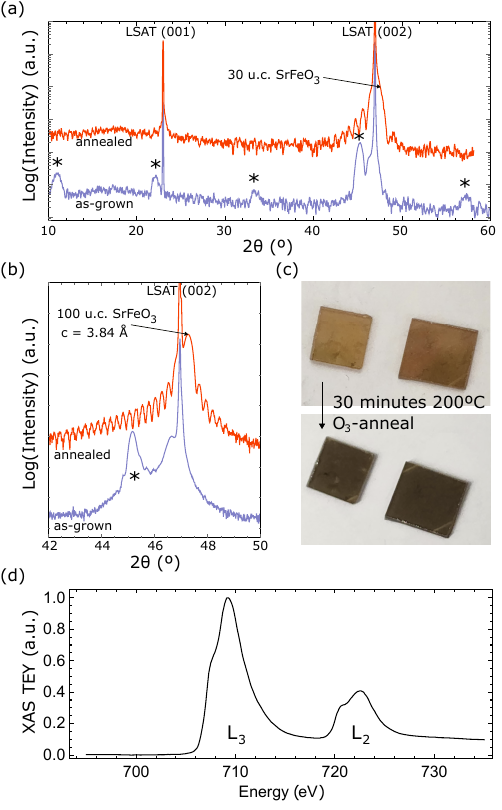}
\caption{\label{xrd}(a) $\theta-2\theta$ x-ray diffraction scans of a 30 u.c. (11 nm) \sfox{} thin film with a 4 u.c. (1.6 nm) capping layer of \sto{} before (blue) and after (red) ozone anneal. (b) $\theta-2\theta$ scans before and after annealing for a 100 u.c. (38 nm) sample with a 3 u.c. (1.2 nm) capping layer of \sto{}. Asterisks indicate the peaks characteristic of the brownmillerite phase. (c) Photograph of two \sfox{} samples before and after ozone-annealing. (d) X-ray absorption in total electron yield around the Fe L$_{2,3}$ edges of a fully oxidized \sfo{} film.} 
\end{figure}

\begin{figure*}
\includegraphics{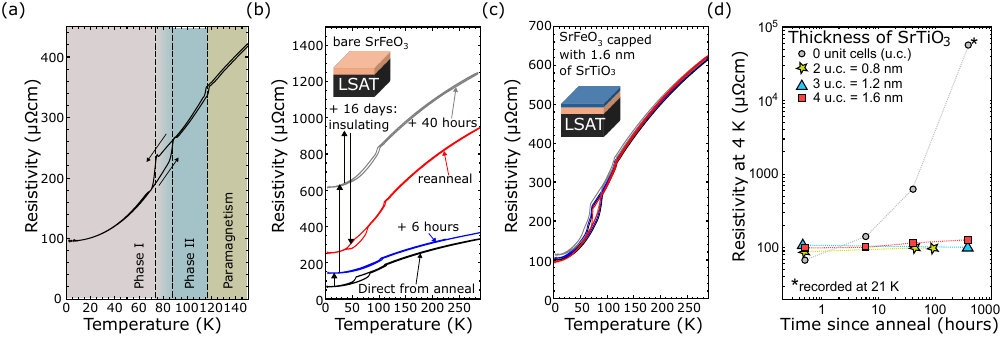}%
\caption{\label{transport}(a) Low temperature close-up of the resistive transitions of \sfo{}, labeled in accordance with Ishiwata \textit{et al} \cite{Ishiwata2011}. (b) Resistivity as a function of temperature showing degradation of the metallicity of bare \sfo{} over several hours. Metallicity is returned after a repeat ozone anneal. (c) With a thin \sto{} capping layer, the metallicity is retained over many hours. The four resistivity versus temperature curves, with the same color coding as in panel (b), are almost indistinguishable. (d) Log-log plot of the time dependence, over several weeks, of the 4 K resistivity of \sfo{} with different thicknesses of capping layer. Dashed lines are guides to the eye. The asterisk marks one data point that was recorded at 21 K due to the resistivity exceeding the measurement limit at lower temperatures.}
\end{figure*}

Epitaxial thin films of strontium ferrite were grown on pseudocubic perovskite (LaAlO$_3$)$_{0.3}$(Sr$_2$AlTaO$_6$)$_{0.7}$ (LSAT) single crystal substrates terminated on the (001) plane \cite{Mateika1991}. When used, an epitaxial capping layer of \sto{} was grown on top. LSAT, with a lattice parameter of 3.87 Å, provides a tensile strain of +0.5 \% to perovskite \sfo{}, which has a bulk lattice parameter of 3.85 Å \cite{MacChesney1965,Takeda1972,Reehuis2012}. Further details on the sample growth by pulsed laser deposition (PLD) are given in the Methods Section. As grown, the films are brownmillerite \sfobm{} or oxygen-deficient perovskite \sfod{} rather than \sfo{}. Unless otherwise stated, all films were topotactically oxidized by an \textit{ex situ} post-anneal in ozone, described in the Methods section. Fig.~\ref{xrd}(a) and (b) show $\theta-2\theta$ scans of the strontium ferrite films as grown (lower curves in each plot) and after the ozone anneal (upper curves in each plot). In both thin (30 u.c. or 11 nm) and thick (100 u.c. or 38 nm) films, there is a complete phase transformation from brownmillerite (peaks indicated by asterisks) to perovskite. Finite thickness fringes indicate the high quality of the perovskite phase. Fig.~\ref{xrd}(c) is an optical photograph of strontium ferrite samples before (top) and after (bottom) ozone-annealing. The color change is due to the transformation from an insulator to a metal.

Fig.~\ref{xrd}(d) shows the Fe L$_{2,3}$ x-ray absorption edge spectrum recorded in total electron yield for an 11 nm film of \sfo{} after standard ozone annealing. The features observed on both edges are consistent with an iron valence state, or oxidation state, of 4+ \cite{Ikeno2004}.

\subsubsection{Electronic transport}
Fig.~\ref{transport}(a) shows typical resistivity data as a function of temperature for a 12 nm thick film of \sfo{} on LSAT (001) substrate with a 1.6 nm \sto{} capping layer. The resistive anomalies indicate N\'{e}el transitions between phases of distinct helimagnetic order and have been labeled according to the established convention \cite{Ishiwata2011}. Residual resistivity is around 100 $\mu \Omega cm $ and the residual resistivity ratio is around 6. Based on these quantifications of metallicity, we suggest that our \sfo{} is the highest quality reported to date among all growth methods (single crystal, powder, molecular beam epitaxy and pulsed laser deposition) and all \textit{ex} and \textit{in situ} oxidation processes (solid oxidizer, oxygen, and ozone) \cite{MacChesney1965,Hayashi2001,Yamada2002,Ishiwata2011,Chakraverty2013,Enriquez2016,Hong2017,Rogge2019,Onose2020,Wang2020e}. Table S1 in the Supplementary Material (SM) summarizes a comparison with prior reports. The high quality of our samples also manifests in the sharpness of the resistive transitions, particularly the transition between Phase I and II, which is often observed as a broad hysteresis in literature. We suggest that the capping layer enhances the quality of our \sfo{} evidenced in electronic transport.

Consistent with previous reports \cite{Wang2020e, Enriquez2016}, the metallicity of bare \sfo{} degrades rapidly with time. This is shown in Fig.~\ref{transport}(b) for an 11 nm \sfo{} film freshly annealed in ozone. The aging of the sample takes place at room temperature in air. Notably, the black curve evolves to the blue curve after only six hours, at which point the resistivity has doubled and the lower temperature Phase I-Phase II N\'{e}el transition is no longer discernible. After 16 days the transport behavior is insulating and the resistivity approaches 0.1 $\Omega cm$. After a reanneal in ozone, the metallicity can be returned (red curve), supporting the assertion that the degradation is due to oxygen loss. We note that, although the residual resistivity rises upon reannealing (compared to initial anneal), the residual resistivity ratio is quite similar.

Fig.~\ref{transport}(c) displays four resistivity curves, corresponding to the same aging periods as the data in panel (b), but here the 11 nm film of \sfo{} is capped with 1.6 nm of \sto{}. The curves are indistinguishable. Thus, the degradation of the metallicity of \sfo{} can be prevented by the addition of a thin capping layer of \sto{}.

Extended time-dependent low temperature resistivity data are plotted in Fig.~\ref{transport}(d) where it is clear that even a \sto{} capping layer of less than 1 nm in thickness is sufficient to maintain the metallicity of \sfo{} over weeks.

Previous work by Enriquez \textit{et al} has shown that a 300 nm capping layer of amorphous \alumina{} maintains a constant resistivity of already oxygen-deficient \sfod{} over several days \cite{Enriquez2016}. Our result is a significant improvement for two reasons. First, the ex situ ozone anneal can be performed with the \sto{} cap because oxygen transport through the heterostructure appears sufficiently rapid at the $\approx$ 200 $^\circ$C temperatures at which the annealing takes place to allow oxidation of the underlying \sfox{}, while oxygen loss is strongly suppressed at room temperature. We speculate that this reflects a temperature-dependent kinetic barrier that may qualitatively function as a “valve”, allowing us to achieve the properties consistent with fully oxidized \sfo{} shown in Fig.~\ref{transport}(a). On the other hand, the 300 nm \alumina{} cap employed by Enriquez et al may provide a larger kinetic barrier against complete initial oxidation. Second, a capping layer of less than 1 nm in thickness does not impede measurement, with the exception of only the most surface-sensitive techniques.

Like Enriquez and coworkers, we believe that the observed degradation of metallicity in \sfo{} is due to oxygen loss as the iron strives to return to the energetically more favorable Fe$^{3+}$ valence state. The effect of the capping layer and the renewed metallicity upon reannealing are our first pieces of evidence.

\begin{figure}
\includegraphics[width=\columnwidth]{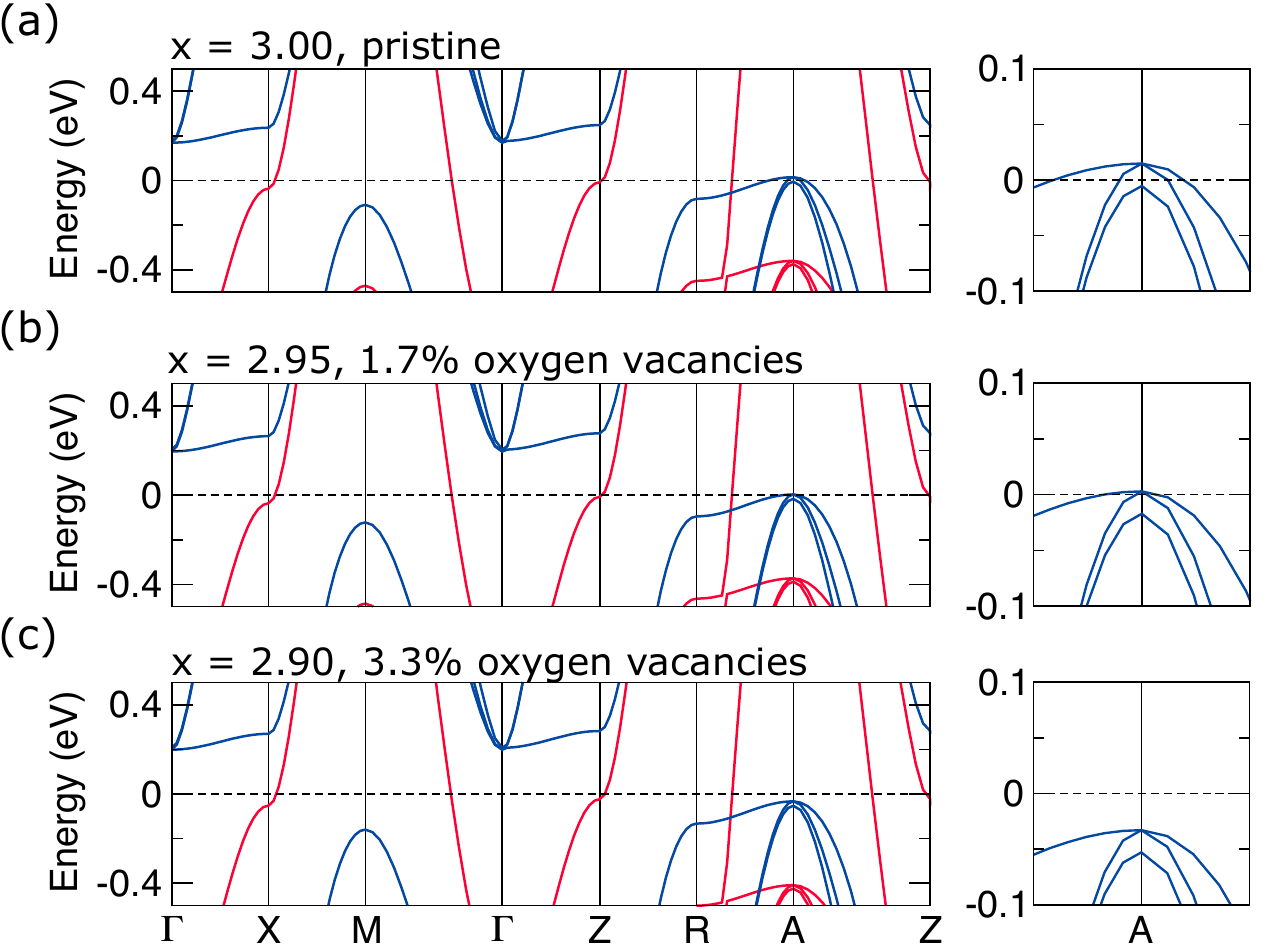}
\caption{\label{DFT}Computed band structure of \sfox{} with x = 3.0 (a), x = 2.95 (b) and x = 2.9 (c). Red and blue bands denote spin up and spin down respectively. Panels on the right display a close-up view around the Fermi level at the A-point.}
\end{figure}

\subsubsection{First principles calculations}
To gain further insight into the high sensitivity of conductivity to oxygen loss we carry out first principles density functional theory (DFT) calculations.
The band structure of pristine \sfo{} reveals three hole-like bands at the A point with a finite density of states at the Fermi level, see Fig. \ref{DFT}(a). Upon increasing oxygen vacancies to 1.7\%, these bands shift to lower energy until their band maximum aligns with the Fermi level, see Fig. \ref{DFT}(b). At 3.3\% oxygen vacancies, these bands no longer intersect with the Fermi level and any carriers they contribute are frozen out. With on order 1\% oxygen vacancies we, therefore, expect a significant decrease in electronic conductivity, consistent with our resistivity data shown in Fig. \ref{transport}). While these calculations do not provide a direct determination of oxygen content, they suggest that relatively small deviations from stoichiometry are a plausible mechanism for the observed transport evolution.

We note that X-ray diffraction does not show a significant change in the film peak position in 2$\theta$ over the course of the metallicity degradation, as shown in the SM Fig.\ S1. This may be because oxygen loss on the scale modeled by our DFT is too small to result in a significant expansion of the lattice parameter. It may also be because \sfo{} is metallic and is able to locally screen the distribution of oxygen vacancies.

Our first principles calculations lend support to our conclusion that oxygen loss is the source of the degradation of metallicity of \sfo{} and, furthermore, that the off-stoichiometry may only be on the order of one percent.

\subsubsection{Transmission electron microscopy}

\begin{figure*}
\includegraphics[width=17.5cm]{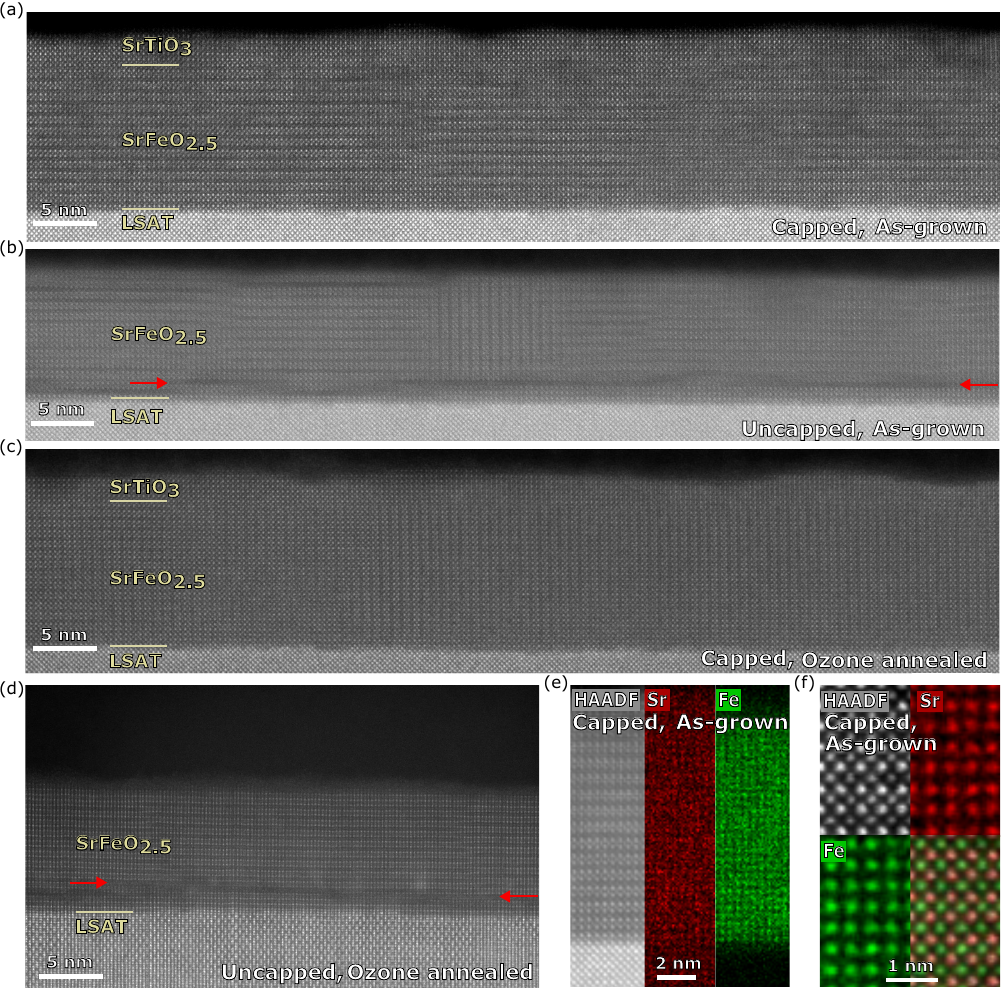}
\caption{\label{stem}(a) Low magnification high angle annular dark field (HAADF) image of an 11 nm thick film of as-grown \sfobm{} with a 1.6 nm \sto{} cap. (b) Low magnification HAADF image of an 11 nm as-grown \sfobm{} film with no capping layer. One horizontal fault in the cation sublattice can be identified and is marked with red arrows. (c) Low magnification HAADF image of an 11 nm thick film of \sfobm{} with a 1.6 nm \sto{} capping layer after having been ozone annealed. (d) Low magnification ADF image of the same film as (b) after having been ozone annealed. The cation fault is indicated by red arrows. Both ozone annealed samples in (c) and (d) were initially \sfo{} and reduced unintentionally by the ion beam or electron beam to \sfobm{}. In all HAADF images the chemical interfaces are marked by yellow bars. (e) On the left a high magnification image of the same sample as in (a) with the electron energy loss spectroscopy (EELS) signal of strontium (middle) and iron (right) taken from the same region. (f) Energy-dispersive x-ray spectroscopy (EDX) maps in a small region of the same sample as in (a) and (e). The combined EDX signals of strontium and iron is shown in the lower right panel.}
\end{figure*}

Recently, Wang and coworkers suggest that the metallicity degradation in \sfo{} has a contribution from extended defects in the cation sublattice that they describe as strontium-rich nanogaps \cite{Wang2020e}. They observe these nanogaps in \sfo{} grown on different substrates, including LSAT. The nanogaps, which they further suggest are a strain relaxation mechanism, are not observed in our samples, as evidenced by our scanning transmission electron microscopy (STEM) analysis.

Fig.~\ref{stem} shows a series of low magnification high angle annular dark field (HAADF) images. Note that ozone annealing fully oxidizes the films to produce perovskite \sfo{}, as evidenced by x-ray diffraction, but either the focused ion beam to prepare the STEM lamella, or the electron beam in the microscope, further reduce the film back to brownmillerite. This is a known challenge in \sfo{} \cite{Lebedev2004} that has previously been successfully mitigated by alternative experimental strategies, such as cryogenic focused ion beam lamella preparation \cite{Ferreiro-Vila2019}. Reduction of \sfobm{} to the infinite-layer \sfoil{} has also been observed under the electron beam during STEM imaging, as we show in SM Fig. S2, and has been been reported by others \cite{Xing2024}. Therefore, beam-induced oxygen loss, as observed in many oxides \cite{Jiang2016a}, is not unprecedented in strontium ferrite.

Fig.~\ref{stem} (a) shows a STEM image of a standard, 11 nm as-grown (before ozone anneal) \sfobm{} sample with a 1.6 nm \sto{} capping layer. The dark bands in the image represent the oxygen vacancy channels of the brownmillerite \sfobm{}. Crucially, the strontium-iron cation sublattice is an interpenetrating primitive cubic lattice with negligible extended defects.

To address the possibility that the addition of the \sto{} capping layer prevents the formation of extended cation defects, we also perform STEM analyses on an uncapped as-grown \sfobm{} sample, as shown in Fig.~\ref{stem} (b). As before, the narrow dark bands in the HAADF image are the oxygen vacancy channels of the brownmillerite structure. There is an additional, wider, dark band close to the substrate (indicated by the red arrows on either end). This is a rocksalt-like antiphase cation defect, yet distinct from the denser and more significant extended defects observed by Wang \textit{et al} \cite{Wang2020e}.

To exclude the possibility that Sr-rich nanogaps form only during the oxidizing process, we perform STEM imaging on \textit{annealed} samples that were, before annealing, equivalent to those shown in Fig.~\ref{stem}(a) and (b). In the capped sample (c) no extended defects are observed and the cation sublattice is pristine. In the uncapped sample (d) we observe the same extended defect near the interface as in the as-grown counterpart but no other defects. 

We also note that regions of vertical oxygen vacancy channels are observed in many samples, with and without capping layer. The coexistence of both orientations of brownmillerite structure, as well as the low energy barrier to switch between them, has been observed previously \cite{Khare2018}.

Finally, Fig.~\ref{stem} (e) and (f) display the electron energy loss spectroscopy (EELS) and energy dispersive x-ray spectroscopy (EDX) maps, respectively, for the same as-grown sample as in Fig.~\ref{stem}(a). With this element specificity, we have further verified that the cation sublattice is close to pristine in this sample. 

We therefore suggest that, while the degradation of metallicity in \sfo{} without a capping layer is ubiquitous, the development of extended cation defects, e.g., strontium-rich nanogaps, is not. While the \sfo{} films are not defect-free, particularly when there is no capping layer, the relative sparseness of the defects, as well as their general orientation in the plane, cannot explain the increase in resistivity of orders of magnitude that we observe in Fig.~\ref{transport}(b).

\section{Discussion and conclusion}
We have demonstrated that the degraded metallicity of \sfo{} can be explained by oxygen stoichiometry only. We infer that the oxygen loss is relatively low, although the absolute value is not directly measured in this work (direct quantification of oxygen loss at these levels remains challenging in thin films). The absence of detectable shifts in lattice parameter, together with DFT calculations showing strong sensitivity of the electronic structure to low vacancy concentrations, suggests that oxygen loss on the order of a few percent may be sufficient to induce insulating behavior.

We believe that the strong negative charge transfer character of \sfo{} plays an important role \cite{Green2024}. Nominally d$^4$, a significant portion of the oxygen 2$p$ electrons are transferred to the iron, making the electronic configuration closer to d$^5$L where L denotes a ligand hole \cite{Rogge2019,Rogge2019b,Takegami2024}. The relevance of oxygen states at the Fermi level leads to the strong sensitivity of resistivity to oxygen content.

Further supporting our conclusion that oxygen loss alone is responsible for degraded metallicity in \sfo{}, our STEM imaging shows a close-to-pristine cation sublattice with and without a capping layer, and with and without ozone annealing.

As a materials synthesis strategy, we have shown that a thin, less than 1 nm, capping layer of band insulator \sto{} permits ozone oxidation of the underlying ferrite at elevated temperatures while retaining metallicity over several weeks at room temperature. As shown in SM Fig. S3, our capped \sfo{} does not exhibit significant degradation until around 100 ºC. Previous work on SrVO$_3$, which has a tendency to gain oxygen, has found amorphous capping layers of 4 - 10 nm thickness to be effective in retaining the stoichiometry \cite{Caspi2022,Mezhoud2025}. A 6 u.c. (2.4 nm) crystalline \sto{} capping layer has previously shown potential in stabilizing Mo$^{4+}$ molybdates over 5 minutes up to 450 ºC \cite{Salg2020}. Those authors claim that the continuity of the transition metal valence at the interface with \sto{} (Ti$^{4+}$--Mo$^{4+}$) may help the stability, which would also be true in the case of \sfo{} (Ti$^{4+}$--Fe$^{4+}$). We propose that the stability of our system is further enhanced by the compressive strain of -0.9\% that the \sto{} experiences due to the epitaxial relationship with the LSAT substrate, transmitted through the \sfo{} film. There is evidence that tensile strain promotes oxygen diffusion in perovskite oxides while compressive strain suppresses it \cite{Kubicek2013, Iglesias2018}. Amorphous capping layers and tensile strained or relaxed crystalline capping layers may, therefore, be less effective than the compressively-strained crystalline capping layer of \sto{} that we use here. These observations are consistent with a temperature-dependent kinetic barrier to oxygen transport. However, the present measurements do not distinguish whether the dominant limitation arises from oxygen surface exchange, interface kinetics, or bulk diffusion through the capping layer.

We also show that metallicity can be returned with a repeat ozone anneal. This will enable more intensive, multi-modal measurements of \sfo{}, and may be generalizable to other oxides with an unfavorable valence state.\\

\section{Methods}
\subsubsection*{Sample preparation}
Thin film samples were grown by pulsed laser deposition by a KrF excimer laser of 248 nm wavelength operating at 3 Hz. The laser fluence was calculated to be 0.97 J cm$^{-2}$. The target was a pressed powder of \sfobm{}. All samples reported in this work were grown in 50 mTorr of flowing oxygen at 620 $^\circ$C, as measured by a pyrometer detecting the front side of the substrate. \sto{} capping layers were grown \textit{in situ} at the same conditions as, and immediately following, \sfobm{}. A single crystal of \sto{} served as the target. Warm up and cool down were ramps of 100 $^\circ$C per minute and were carried out in the same oxygen partial pressure as the growth. \textit{Ex situ} ozone anneal was performed at 180 - 260 $^\circ$C for 0.5 to 3 hours in a UV-assisted ozone plasma generator with an ozone concentration of approximately 5000 ppm.

\subsubsection*{X-ray diffraction}
$\theta$-2$\theta$ measurements were acquired on a PANalytical X'Pert 2 Pro. 

\subsubsection*{X-ray absorption}
X-ray absorption measurements were performed at beamline 10-1 at Stanford Synchrotron Radiation Lightsource. Data were taken in total electron yield mode with 45-degree x-ray incidence and 1 mm$^2$ beam size.

\subsubsection*{Transport measurements}
Gold contacts of 30 nm thickness were evaporated onto the samples in van der Pauw geometry then aluminum wires were ultrasonically bonded. 4-point transport measurements were carried out in a $^4$He cryostat.

\subsubsection*{Scanning transmission electron microscopy}
TEM specimen lamellae from as-grown samples (Fig. \ref{stem}(a, b, e, f)) were prepared using a focused ion beam (FIB) Helios 5 UX.
Scanning transmission electron microscopy (STEM) images of as-grown samples (Fig. \ref{stem}(a, b)) were recorded using a double-corrected Thermo Fisher SPECTRA 300 operated at 200 kV using a convergence semi-angle of 20 mrad and a beam current of 50 pA. Energy loss spectrum images were acquired using a continuum spectrometer equipped with a direct electron detection K3 camera from Gatan. A four quadrant Super-X windowless silicon drift detector system was used to acquire the energy dispersive x-ray spectroscopy (EDX) signal. The EDX maps were generated using the Thermo Fisher Scientific Velox software following the Cliff-Lorimer approach with Brown-Powellxf ionization cross-section models.

STEM characterization of a capped, ozone-annealed sample (Fig. \ref{stem}(c)) was performed on a cross-sectional lamella prepared with the standard FIB lift-out procedure using a Thermo Fisher Helios G4 UX FIB. ADF-STEM imaging was performed on aberration-corrected TFS Titan Themis 300 operating at 300 kV with probe convergence semi-angle of 30 mrad.

STEM samples of the oxygen annealed uncapped sample were prepared by milling out and mounting a lamella onto a copper grid using a Thermo Fisher Scientific Helios 5 CX Dual Beam FIB scanning electron microscope (SEM), with gallium as the source. The lamella was then thinned using a Thermo Scientific Helios 5 Hydra UX Dual Beam plasma FIB to mitigate damage caused by the harsher gallium-ion beam.
High-angle annular dark field images were acquired on a JEOL Neo-ARM operating at 200 kV with a nominally 30 mrad convergence semi-angle and 16 $\mu$s dwell time. To minimize beam damage to the region of interest the electron beam was first focused on the sample in a sacrificial region of the substrate, the stage was then moved to the film and immediately blanked to allow for the stage to settle without damaging the sample. The beam was then un-blanked at a high magnification with a small focus region selected, which was rapidly focused, and then the magnification decreased, which minimized the region of the sample that incurred significant damage. Images with increased magnification were then acquired.

\subsubsection*{Density functional theory}
We perform first-principles density functional calculations within the Perdew–Burke–Ernzerhof exchange–correlation functional revised for solids (PBEsol) \cite{PBEsol:2008} as implemented in the Vienna {\it Ab initio} Simulation Package ({\sc vasp}) with the projector augmented wave (PAW) method \cite{kresse_ab_1993, kresse_efficiency_1996, kresse_efficient_1996, perdew_generalized_1996, blochl_projector_1994, kresse_ultrasoft_1999}.  The core and valence electrons are treated using the following  electronic configurations:
4$s^{2}$4p$^6$5s$^2$ (Sr), 3d$^7$4s$^1$ (Fe), 2s$^2$2p$^4$ (O), and a 600~eV plane wave cutoff.
Calculations were carried out using the experimental lattice parameters, while internal distortions were neglected since none are observed experimentally. 
To account for the electronic consequences of oxygen vacancies, we adopt a background-charge approach that varies the total electron number, effectively simulating carrier doping with a uniform compensating background to ensure charge neutrality.

\section*{Supplementary Material}
A Supplementary Material (SM) file is available. The file contains a quantitative comparison of metallicity between \sfo{} samples reported in literature, evidence that the c-axis lattice parameter is not significantly affected by low levels of oxygen loss, further STEM images demonstrating the reduction to infinite-layer \sfoil{} and an experiment upon heating capped \sfox{} in air on a hot plate suggesting that oxygen loss is insignificant up to 100 ºC.

\begin{acknowledgments}
The work at SLAC/Stanford was supported by the US Department of Energy, Office of Basic Energy Sciences, Division of Materials Sciences and Engineering, under contract no. DE-AC02-76SF00515. Part of this work was performed at the Stanford Nano Shared Facilities (SNSF) RRID:SCR\_023230, supported by the National Science Foundation under award ECCS-2026822.
Authors acknowledge the use of instrumentation as well as the technical advice provided by the Joint Electron Microscopy Center at ALBA (JEMCA) and funding from Grant IU16-014206 (METCAM-FIB) to ICN2 funded by the European Union through the European Regional Development Fund (ERDF), with the support of the Ministry of Research and Universities, Generalitat de Catalunya. Authors acknowledge F. Belarre and M. Rosado from ICN2 for the FIB lamellae preparation of the as-grown samples.
STEM HAADF imaging using the Neo-ARM was supported by the Center for Nanophase Materials Sciences, (CNMS), which is a DOE Office of Science User Facility using instrumentation within ORNL’s Materials Characterization Core provided by UT-Battelle, LLC, under Contract No. DE-AC05-00OR22725 with the DOE. 
W.J.K. acknowledges funding by the 2024 BK21 FOUR Program of Pusan National University and National Research Foundation (NRF) of Korea (No. RS-2024-00404737).
LB acknowledges Dr. Jinkwon Kim's contribution in reoxidizing the samples for electron microscopy measurements at Cornell.
The authors acknowledge the support and advice of the late Professor Lena F. Kourkoutis to this study. 
\end{acknowledgments}

\section*{Data Availability Statement}

The data that support the findings of this study are available in Northwestern University's ARCH Repository, https://doi.org/10.21985/n2-emy6-vt31.

\bibliography{PAPERS-SFO, puggioni}

@article{Ferreiro-Vila2019,
author = {Ferreiro-Vila, El{\'{i}}as and Blanco-Canosa, Santiago and {Lucas del Pozo}, Irene and Vasili, Hari Babu and Mag{\'{e}}n, C{\'{e}}sar and Ibarra, Alfonso and Rubio-Zuazo, Juan and Castro, Germ{\'{a}}n R. and Morell{\'{o}}n, Luis and Rivadulla, Francisco},
journal = {Advanced Functional Materials},
pages = {1901984},
title = {{Room-Temperature AFM Electric-Field-Induced Topotactic Transformation between Perovskite and Brownmillerite SrFeO$_x$ with Sub-Micrometer Spatial Resolution}},
volume = {29},
year = {2019}
}

@article{Xing2024,
author = {Xing, Yaolong and Kim, Inhwan and Kang, Kyeong Tae and Byun, Jinho and Choi, Woo Seok and Lee, Jaekwang and Oh, Sang Ho},
isbn = {4155702401},
issn = {17554349},
journal = {Nature Chemistry},
pages = {66},
publisher = {Springer US},
title = {{Monitoring the formation of infinite-layer transition metal oxides through in situ atomic-resolution electron microscopy}},
volume = {17},
year = {2024}
}

@article{Khare2018,
author = {Khare, Amit and Lee, Jaekwang and Park, Jaeseoung and Kim, Gi Yeop and Choi, Si Young and Katase, Takayoshi and Roh, Seulki and Yoo, Tae Sup and Hwang, Jungseek and Ohta, Hiromichi and Son, Junwoo and Choi, Woo Seok},
issn = {19448252},
journal = {ACS Applied Materials and Interfaces},
number = {5},
pages = {4831--4837},
pmid = {29327588},
title = {{Directing Oxygen Vacancy Channels in SrFeO$_{2.5}$ Epitaxial Thin Films}},
volume = {10},
year = {2018}
}

@article{Kubicek2013,
author = {Kubicek, Markus and Cai, Zhuhua and Ma, Wen and Yildiz, Bilge and Hutter, Herbert and Fleig, J{\"{u}}rgen},
journal = {ACS Nano},
pages = {3276--3286},
title = {{Tensile Lattice Strain Accelerates Oxygen Surface Exchange and diffusion in La$_{1-x}$Sr$_x$CoO$_{3-\delta}$ thin films}},
volume = {7},
year = {2013}
}

@article{Rogge2019b,
author = {Rogge, Paul C and Shafer, Padraic and Fabbris, Gilberto and Hu, Wen and Arenholz, Elke and Karapetrova, Evguenia and Dean, Mark P M and Green, Robert J and May, Steven J},
journal = {Advanced Materials},
pages = {1902364},
title = {{Depth-Resolved Modulation of Metal – Oxygen Hybridization and Orbital Polarization across Correlated Oxide Interfaces}},
volume = {31},
year = {2019}
}

@article{Onose2020,
author = {Onose, M. and Takahashi, H. and Sagayama, H. and Yamasaki, Y. and Ishiwata, S.},
issn = {24759953},
journal = {Physical Review Materials},
number = {11},
pages = {114420},
publisher = {American Physical Society},
title = {{Complete phase diagram of Sr$_{1-x}$La$_x$FeO$_3$ with versatile magnetic and charge ordering}},
volume = {4},
year = {2020}
}

@article{Fowlie2026,
author = {Fowlie, Jennifer and Li, Jiarui and Puggioni, Danilo and Barreto, Lucas and Yuan, Lin Ding and Rondinelli, James M. and Sutarto, Ronny and Boyko, Teak D. and Orlandi, Fabio and Manuel, Pascal and Khalyavin, Dmitry and Lomeli, Eder G. and Moritz, Brian and Devereaux, Thomas P. and Koroluk, Skylar and Green, Robert J. and May, Steven J. and Hwang, Harold Y.},
journal = {arXiv cond-mat},
pages = {2602.10372v2},
title = {{Biaxial Strain Control of Helimagnetism via Chemical Expansion in Thin Film SrFeO3}},
year = {2026}
}

@article{Ge2019,
author = {Ge, Chen and Liu, Chang-xiang and Zhou, Qing-li and Zhang, Qing-hua and Du, Jian-yu and Li, Jian-kun and Wang, Can and Gu, Lin and Yang, Guo-zhen and Jin, Kui-juan},
issn = {15214095},
journal = {Advanced Materials},
pages = {1900379},
pmid = {30924206},
title = {{A Ferrite Synaptic Transistor with Topotactic Transformation}},
volume = {31},
year = {2019}
}

@article{Ikeno2004,
author = {Ikeno, Hidekazu and Tanaka, Isao and Miyamae, Toru and Mishima, Takahiro and Adachi, Hirohiko and Ogasawara, Kazuyoshi},
issn = {13459678},
journal = {Materials Transactions},
number = {5},
pages = {1414--1418},
title = {{First principles calculation of Fe L$_{2,3}$-edge X-ray absorption near edge structures of iron oxides}},
volume = {45},
year = {2004}
}

@article{Jiang2016a,
author = {Jiang, Nan},
journal = {Reports on Progress in Physics},
pages = {016501},
publisher = {IOP Publishing},
title = {{Electron beam damage in oxides : a review}},
volume = {79},
year = {2016}
}

@article{Nallagatla2019,
author = {Nallagatla, Venkata R. and Heisig, Thomas and Baeumer, Christoph and Feyer, Vitaliy and Jugovac, Matteo and Zamborlini, Giovanni and Schneider, Claus M. and Waser, Rainer and Kim, Miyoung and Jung, Chang Uk and Dittmann, Regina},
issn = {15214095},
journal = {Advanced Materials},
pages = {1903391},
pmid = {31441160},
title = {{Topotactic Phase Transition Driving Memristive Behavior}},
volume = {31},
year = {2019}
}

@article{Caspi2022,
author = {Caspi, Shaked and Shoham, Lishai and Baskin, Maria and Weinfeld, Kamira and Piamonteze, Cinthia and Stoerzinger, Kelsey A. and Kornblum, Lior},
journal = {Journal of Vacuum Science and Technology A: Vacuum, Surfaces, and Films},
pages = {013208},
publisher = {American Vacuum Society},
title = {{The effect of capping layers on the near-surface region of SrVO$_3$ films}},
volume = {40},
year = {2022}
}

@article{Hong2017,
author = {Hong, Deshun and Liu, Changjiang and Pearson, John and Bhattacharya, Anand},
issn = {00036951},
journal = {Applied Physics Letters},
number = {23},
pages = {232408},
title = {{Epitaxial growth of high quality SrFeO$_3$ films on (001) oriented (LaAlO$_3$)$_{0.3}$(Sr$_2$TaAlO$_6$)$_{0.7}$}},
volume = {111},
year = {2017}
}

@article{Ishiwata2020,
author = {Ishiwata, S. and Nakajima, T. and Kim, J. H. and Inosov, D. S. and Kanazawa, N. and White, J. S. and Gavilano, J. L. and Georgii, R. and Seemann, K. M. and Brandl, G. and Manuel, P. and Khalyavin, D. D. and Seki, S. and Tokunaga, Y. and Kinoshita, M. and Long, Y. W. and Kaneko, Y. and Taguchi, Y. and Arima, T. and Keimer, B. and Tokura, Y.},
journal = {Physical Review B},
number = {13},
pages = {134406},
publisher = {American Physical Society},
title = {{Emergent topological spin structures in the centrosymmetric cubic perovskite SrFeO$_3$}},
volume = {101},
year = {2020}
}

@article{Yamada2002,
author = {Yamada, Hiroyuki and Kawasaki, M. and Tokura, Y.},
issn = {00036951},
journal = {Applied Physics Letters},
number = {4},
pages = {622--624},
title = {{Epitaxial growth and valence control of strained perovskite SrFeO$_3$ films}},
volume = {80},
year = {2002}
}

@article{Rogge2019,
author = {Rogge, Paul C. and Green, Robert J. and Sutarto, Ronny and May, Steven J.},
issn = {24759953},
journal = {Physical Review Materials},
pages = {084404},
publisher = {American Physical Society},
title = {{Itinerancy-dependent noncollinear spin textures in SrFeO$_3$, CaFeO$_3$, and CaFeO$_3$/SrFeO$_3$ heterostructures probed via resonant X-ray scattering}},
volume = {3},
year = {2019}
}

@article{Wang2020e,
author = {Wang, Le and Yang, Zhenzhong and Wu, Jinpeng and Bowden, Mark E. and Yang, Wanli and Qiao, Amy and Du, Yingge},
issn = {23972106},
journal = {npj Materials Degradation},
number = {16},
pages = {1--6},
publisher = {Springer US},
title = {{Time- and strain-dependent nanoscale structural degradation in phase change epitaxial strontium ferrite films}},
volume = {4},
year = {2020}
}

@article{Psaroudaki2023,
author = {Psaroudaki, Christina and Peraticos, Elias and Panagopoulos, Christos},
journal = {Applied Physics Letters},
number = {26},
pages = {260501},
publisher = {AIP Publishing LLC},
title = {{Skyrmion qubits: Challenges for future quantum computing applications}},
volume = {123},
year = {2023}
}

@article{Enriquez2016,
author = {Enriquez, Erik and Chen, Aiping and Harrell, Zach and L{\"{u}}, Xujie and Dowden, Paul and Koskelo, Nicholas and Janoschek, Marc and Chen, Chonglin and Jia, Quanxi},
issn = {00036951},
journal = {Applied Physics Letters},
number = {14},
pages = {141906},
title = {{Oxygen vacancy-driven evolution of structural and electrical properties in SrFeO$_{3-\delta}$ thin films and a method of stabilization}},
volume = {109},
year = {2016}
}

@article{Islam2023,
author = {Islam, Rabiul and Li, Peng and Beg, Marijan and Sachdev, Manoj and Miao, Guo Xing},
journal = {Applied Physics Letters},
number = {15},
pages = {152407},
publisher = {AIP Publishing LLC},
title = {{Helimagnet-based nonvolatile multi-bit memory units}},
volume = {122},
year = {2023}
}

@article{Ishiwata2011,
author = {Ishiwata, S. and Tokunaga, M. and Kaneko, Y. and Okuyama, D. and Tokunaga, Y. and Wakimoto, S. and Kakurai, K. and Arima, T. and Taguchi, Y. and Tokura, Y.},
issn = {10980121},
journal = {Physical Review B},
number = {5},
pages = {054427},
title = {{Versatile helimagnetic phases under magnetic fields in cubic perovskite SrFeO$_3$}},
volume = {84},
year = {2011}
}

@article{Takeda1972,
author = {Takeda, Takayoshi and Yamaguchi, Yasuo and Watanabe, Hiroshi},
journal = {Journal of the Physical Society of Japan},
number = {4},
pages = {967},
title = {{Magnetic Structure of SrFeO$_3$}},
volume = {33},
year = {1972}
}

@article{Mezhoud2025,
author = {Mezhoud, Moussa and Rath, Martando and Gascoin, St{\'{e}}phanie and Duprey, Sylvain and Marie, Philippe and Cardin, Julien and Labb{\'{e}}, Christophe and Prellier, Wilfrid and L{\"{u}}ders, Ulrike},
journal = {Nanoscale},
pages = {15319--15330},
publisher = {Royal Society of Chemistry},
title = {{Improving the long-term stability of new- generation perovskite-based TCOs using binary and ternary oxides capping layers}},
volume = {17},
year = {2025}
}

@article{Salg2020,
author = {Salg, Patrick and Zeinar, Lukas and Radetinac, Aldin and Walk, Dominik and Maune, Holger and Jakoby, Rolf and Alff, Lambert and Komissinskiy, Philipp},
journal = {Journal of Applied Physics},
number = {August},
pages = {065302},
publisher = {AIP Publishing LLC},
title = {{Oxygen diffusion barriers for epitaxial thin-film heterostructures with highly conducting SrMoO$_3$ electrodes}},
volume = {127},
year = {2020}
}

@article{Chakraverty2013,
author = {Chakraverty, S. and Matsuda, T. and Wadati, H. and Okamoto, J. and Yamasaki, Y. and Nakao, H. and Murakami, Y. and Ishiwata, S. and Kawasaki, M. and Taguchi, Y. and Tokura, Y. and Hwang, H. Y.},
issn = {10980121},
journal = {Physical Review B},
number = {22},
pages = {220405(R)},
title = {{Multiple helimagnetic phases and topological Hall effect in epitaxial thin films of pristine and Co-doped SrFeO$_3$}},
volume = {88},
year = {2013}
}

@article{Takegami2024,
author = {Takegami, D. and Nakamura, M. and Melendez-Sans, A. and Fujinuma, K. and Nakamura, R. and Yoshimura, M. and Tsuei, K. D. and Tanaka, A. and Gen, M. and Tokunaga, Y. and Ishiwata, S. and Mizokawa, T.},
issn = {24699969},
journal = {Physical Review B},
pages = {235138},
publisher = {American Physical Society},
title = {{Negative charge-transfer energy in SrFeO$_3$ revisited with hard x-ray photoemission spectroscopy}},
volume = {109},
year = {2024}
}

@article{MacChesney1965,
author = {MacChesney, J. B. and Sherwood, R. C. and Potter, J. F.},
doi = {10.1063/1.1697052},
issn = {00219606},
journal = {The Journal of Chemical Physics},
number = {6},
pages = {1907--1913},
title = {{Electric and magnetic properties of the strontium ferrates}},
volume = {43},
year = {1965}
}

@article{Hayashi2001,
author = {Hayashi, Naoaki and Terashima, Takahito and Takano, Mikio},
doi = {10.1039/b103259n},
issn = {13645501},
journal = {Journal of Materials Chemistry},
number = {9},
pages = {2235--2237},
title = {{Oxygen-holes creating different electronic phases in Fe$^{4+}$-oxides: Successful growth of single crystalline films of SrFeO$_3$ and related perovskites at low oxygen pressure}},
volume = {11},
year = {2001}
}

@article{Mateika1991,
author = {Mateika, D. and Kohler, H. and Laudan, H. and V{\"{o}}lkel, E.},
journal = {Journal of Crystal Growth},
number = {1991},
pages = {447--456},
title = {{Mixed-perovskite substrates for high-T$_c$ superconductors}},
volume = {109},
year = {1991}
}

@article{Iglesias2018,
author = {Iglesias, Lucia and Gomez, Andres and Gich, Marti and Rivadulla, Francisco},
doi = {10.1021/acsami.8b12019},
journal = {ACS Applied Materials and Interfaces},
pages = {35367},
title = {{Tuning Oxygen Vacancy Diffusion through Strain in SrTiO$_3$ Thin Films}},
volume = {10},
year = {2018}
}

@article{Green2024,
author = {Green, Robert J and Sawatzky, George A.},
journal = {Journal of the Physical Society of Japan},
pages = {121007},
title = {{Negative Charge Transfer Energy in Correlated Compounds}},
volume = {93},
year = {2024}
}

@article{Schmidt2000,
author = {Schmidt, M},
journal = {Journal of Physics and Chemistry of Solids},
pages = {1363--1365},
title = {{Composition adjustment of non-stoichiometric strontium ferrite SrFeO$_{3-\delta}$}},
volume = {61},
year = {2000}
}

@article{Takeda1986,
author = {Takeda, Y and Kanno, K and Takada, T and Yamamoto, O and Takano, M and Nakayama, N and Bando, Y},
journal = {Journal of Solid State Chemistry},
pages = {237--249},
title = {{Phase Relation in the Oxygen Nonstoichiometric System SrFeO$_x$ (2.5 $\leq$ x $\leq$ 3.0)}},
volume = {63},
year = {1986}
}

@article{Lebedev2004,
author = {Lebedev, O I and Verbeeck, J and van Tendeloo, G and Hayashi, N and Terashima, T},
journal = {Philosophical Magazine},
pages = {3825},
title = {{Structure and microstructure of epitaxial Sr$_n$Fe$_n$O$_{3n-1}$ films}},
volume = {84},
year = {2004}
}

@article{Reehuis2012,
author = {Reehuis, M and Ulrich, C and Maljuk, A and Niedermayer, Ch and Ouladdiaf, B and Hoser, A and Hofmann, T and Keimer, B},
journal = {Physical Review B},
pages = {184109},
title = {{Neutron diffraction study of spin and charge ordering in SrFeO$_{3-\delta}$}},
volume = {85},
year = {2012}
}

@STRING( EL = "Europhys. Lett.")

@STRING( PRL = "Physical Review Letters")

@article{kresse_ab_1993,
	title = {Ab initio molecular dynamics for liquid metals},
	volume = {47},
    year =1993,
	url = {https://link.aps.org/doi/10.1103/PhysRevB.47.558},
	doi = {10.1103/PhysRevB.47.558},
	pages = {558--561},
	number = {1},
	journal = {Physical Review B},
	shortjournal = {Phys. Rev. B},
	author = {Kresse, G. and Hafner, J.},
	urldate = {2025-04-09},
	date = {1993-01-01},
}

@article{kresse_efficiency_1996,
	title = {Efficiency of ab-initio total energy calculations for metals and semiconductors using a plane-wave basis set},
	volume = {6},
	issn = {0927-0256},
	doi = {10.1016/0927-0256(96)00008-0},
	pages = {15--50},
	number = {1},
	journal = {Computational Materials Science},
	shortjournal = {Computational Materials Science},
	author = {Kresse, G. and Furthmüller, J.},
	urldate = {2025-04-09},
    year =1996,
	date = {1996-07-01},
}

@article{kresse_efficient_1996,
	title = {Efficient iterative schemes for ab initio total-energy calculations using a plane-wave basis set},
	volume = {54},
	doi = {10.1103/PhysRevB.54.11169},
	pages = {11169--11186},
	number = {16},
	journal = {Physical Review B},
	shortjournal = {Phys. Rev. B},
	author = {Kresse, G. and Furthmüller, J.},
	urldate = {2025-04-09},
	date = {1996-10-15},
    year = 1996,
}

@article{perdew_generalized_1996,
	title = {Generalized Gradient Approximation Made Simple},
	volume = {77},
    year =1996,
	url = {https://link.aps.org/doi/10.1103/PhysRevLett.77.3865},
	doi = {10.1103/PhysRevLett.77.3865},
	pages = {3865--3868},
	number = {18},
	journal = {Physical Review Letters},
	shortjournal = {Phys. Rev. Lett.},
	author = {Perdew, John P. and Burke, Kieron and Ernzerhof, Matthias},
	urldate = {2025-04-09},
	date = {1996-10-28},
}

@article{blochl_projector_1994,
	title = {Projector augmented-wave method},
	volume = {50},
	doi = {10.1103/PhysRevB.50.17953},
	pages = {17953--17979},
	number = {24},
	journal = {Physical Review B},
	shortjournal = {Phys. Rev. B},
	author = {Blöchl, P. E.},
	urldate = {2025-04-09},
	date = {1994-12-15},
	year = 1994,
}

@article{kresse_ultrasoft_1999,
	title = {From ultrasoft pseudopotentials to the projector augmented-wave method},
	volume = {59},
    year =1999,
	doi = {10.1103/PhysRevB.59.1758},
	pages = {1758--1775},
	number = {3},
	journal = {Physical Review B},
	shortjournal = {Phys. Rev. B},
	author = {Kresse, G. and Joubert, D.},
	urldate = {2025-04-09},
	date = {1999-01-15},
}

@Article{PBEsol:2008,
  title = {Restoring the Density-Gradient Expansion for Exchange in Solids and Surfaces},
  author = {Perdew, John P. and Ruzsinszky, Adrienn  and Csonka, G\'abor I. and Vydrov, Oleg A. and Scuseria, Gustavo E. and Constantin, Lucian A. and Zhou, Xiaolan  and Burke, Kieron },
  journal = PRL,
  volume = {100},
  number = {13},
  pages = {136406},
  numpages = {4},
  year = {2008},
}

\end{document}